\documentclass[twocolumn,twoside,slac_two]{revtex4}
\usepackage{graphicx}
\usepackage{fancyhdr}
\newcommand{\vvsmallrule}{\rule[-1.0mm]{0.0cm}{0.6cm}}

\RequirePackage{xspace}

\usepackage{relsize}
\def\babar{\mbox{\slshape B\kern-0.1em{\smaller A}\kern-0.1em
    B\kern-0.1em{\smaller A\kern-0.2em R}}}

\def\en         {\ensuremath{e^-}\xspace}   
\def\ep         {\ensuremath{e^+}\xspace}

\def\g     {\ensuremath{\gamma}\xspace}

\def\s     {\ensuremath{s}\xspace}

\def\b     {\ensuremath{b}\xspace}

\def\piz   {\ensuremath{\pi^0}\xspace}

\def\pip   {\ensuremath{\pi^+}\xspace}
\def\pim   {\ensuremath{\pi^-}\xspace}

\def\Kbar  {\kern 0.2em\overline{\kern -0.2em K}{}\xspace}

\def\Kz    {\ensuremath{K^0}\xspace}
\def\Kzb   {\ensuremath{\Kbar^0}\xspace}
\def\KzKzb {\ensuremath{\Kz \kern -0.16em \Kzb}\xspace}
\def\Kp    {\ensuremath{K^+}\xspace}
\def\Km    {\ensuremath{K^-}\xspace}

\def\KpKm  {\ensuremath{\Kp \kern -0.16em \Km}\xspace}
\def\KS    {\ensuremath{K^0_{\scriptscriptstyle S}}\xspace}

\def\Kstar   {\ensuremath{K^*}\xspace}

\def\Dbar    {\kern 0.2em\overline{\kern -0.2em D}{}\xspace}
\def\Db      {\ensuremath{\Dbar}\xspace}
\def\Dz      {\ensuremath{D^0}\xspace}
\def\Dzb     {\ensuremath{\Dbar^0}\xspace}
\def\DzDzb   {\ensuremath{\Dz {\kern -0.16em \Dzb}}\xspace}
\def\Dp      {\ensuremath{D^+}\xspace}
\def\Dm      {\ensuremath{D^-}\xspace}

\def\DpDm    {\ensuremath{\Dp {\kern -0.16em \Dm}}\xspace}

\def\B       {\ensuremath{B}\xspace}
\def\Bbar    {\kern 0.18em\overline{\kern -0.18em B}{}\xspace}

\def\BB      {\ensuremath{B\Bbar}\xspace} 
\def\Bz      {\ensuremath{B^0}\xspace}
\def\Bzb     {\ensuremath{\Bbar^0}\xspace}
\def\BzBzb   {\ensuremath{\Bz {\kern -0.16em \Bzb}}\xspace}
\def\Bu      {\ensuremath{B^+}\xspace}
\def\Bub     {\ensuremath{B^-}\xspace}
\def\Bp      {\ensuremath{\Bu}\xspace}

\def\BpBm    {\ensuremath{\Bu {\kern -0.16em \Bub}}\xspace}

\def\BorBbar    {\kern 0.18em\optbar{\kern -0.18em B}{}\xspace}
\def\DorDbar    {\kern 0.18em\optbar{\kern -0.18em D}{}\xspace}
\def\KorKbar    {\kern 0.18em\optbar{\kern -0.18em K}{}\xspace}

\mathchardef\Upsilon="7107
\def\Y#1S{\ensuremath{\Upsilon{(#1S)}}\xspace}

\mathchardef\Deltares="7101
\mathchardef\Xi="7104
\mathchardef\Lambda="7103
\mathchardef\Sigma="7106
\mathchardef\Omega="710A

\def\Deltabar{\kern 0.25em\overline{\kern -0.25em \Deltares}{}\xspace}
\def\Lbar{\kern 0.2em\overline{\kern -0.2em\Lambda\kern 0.05em}\kern-0.05em{}\xspace}
\def\Sigbar{\kern 0.2em\overline{\kern -0.2em \Sigma}{}\xspace}
\def\Xibar{\kern 0.2em\overline{\kern -0.2em \Xi}{}\xspace}
\def\Obar{\kern 0.2em\overline{\kern -0.2em \Omega}{}\xspace}
\def\Nbar{\kern 0.2em\overline{\kern -0.2em N}{}\xspace}
\def\Xb{\kern 0.2em\overline{\kern -0.2em X}{}\xspace}

\def\BR         {{\ensuremath{\cal B}\xspace}}

\def\mes        {\mbox{$m_{\rm ES}$}\xspace}

\def\DeltaE     {\mbox{$\Delta E$}\xspace}

\newcommand{\tev}{\ensuremath{\mathrm{\,Te\kern -0.1em V}}\xspace}
\newcommand{\gev}{\ensuremath{\mathrm{\,Ge\kern -0.1em V}}\xspace}
\newcommand{\mev}{\ensuremath{\mathrm{\,Me\kern -0.1em V}}\xspace}
\newcommand{\kev}{\ensuremath{\mathrm{\,ke\kern -0.1em V}}\xspace}
\newcommand{\ev}{\ensuremath{\mathrm{\,e\kern -0.1em V}}\xspace}
\newcommand{\gevc}{\ensuremath{{\mathrm{\,Ge\kern -0.1em V\!/}c}}\xspace}
\newcommand{\mevc}{\ensuremath{{\mathrm{\,Me\kern -0.1em V\!/}c}}\xspace}
\newcommand{\gevcc}{\ensuremath{{\mathrm{\,Ge\kern -0.1em V\!/}c^2}}\xspace}
\newcommand{\mevcc}{\ensuremath{{\mathrm{\,Me\kern -0.1em V\!/}c^2}}\xspace}

\def\mus  {\ensuremath{\rm \,\mus}\xspace}

\def\mus        {\ensuremath{\,\mu{\rm s}}\xspace}    

\def\to                 {\ensuremath{\rightarrow}\xspace}

\def\pep2{PEP-II}

\def\gsim{{~\raise.15em\hbox{$>$}\kern-.85em
          \lower.35em\hbox{$\sim$}~}\xspace}
\def\lsim{{~\raise.15em\hbox{$<$}\kern-.85em
          \lower.35em\hbox{$\sim$}~}\xspace}

\def\Vub  {\ensuremath{|V_{ub}|}\xspace}
\def\Vcb  {\ensuremath{|V_{cb}|}\xspace}

\xspace

\newcommand{\epjBase}        {Eur.\ Phys.\ Jour.\xspace}
\newcommand{\jprlBase}       {Phys.\ Rev.\ Lett.\xspace}
\newcommand{\jprBase}        {Phys.\ Rev.\xspace}
\newcommand{\jplBase}        {Phys.\ Lett.\xspace}

\newcommand{\npBase}         {Nucl.\ Phys.\xspace}

\newcommand{\epjc}      [1]  {\epjBase\ C~{\bf #1}}

\newcommand{\npb}       [1]  {\npBase\ B~{\bf #1}}

\newcommand{\plb}       [1]  {\jplBase\ B~{\bf #1}}

\newcommand{\jprl}      [1]  {\jprlBase\ {\bf #1}}
\newcommand{\jprd}      [1]  {\jprBase\ D~{\bf #1}}

\def\jetset74   {\mbox{\tt Jetset \hspace{-0.5em}7.\hspace{-0.2em}4}\xspace}

 \def\eg         {\ensuremath{E_{\gamma} }\xspace}

 \def\mes        {\ensuremath {M_\mathit{ES}}\xspace}
 
 \def\de         {\ensuremath {\Delta E^{*}}\xspace}

 \def\mupisq    {\ensuremath{\mu_{\pi}^2}\xspace}

 \def\mb         {\ensuremath{m_{b} }\xspace}

 \def\acp        {\ensuremath{A_{CP}}\xspace}

 \def\bxsg       {\ensuremath{B \to X_{s} \gamma}\xspace}

 \def\bkg        {\ensuremath{B \to \Kstar \gamma}\xspace}

\def\aveDelta#1 {\ensuremath{\langle \Delta_{total}#1 \rangle}\xspace}

\def\ecut {\ensuremath{E_{\rm cut}}}
 
 \def\meg {\ensuremath{\langle E_{\gamma} \rangle}}
 
 \def\vegs {\ensuremath{\langle (E_{\gamma} -\meg)^2 \rangle}}

\def\rhoz {\rho^0}
\def\rhop {\rho^+}

\def\mb      {\ensuremath {M_{\mbox{\scriptsize ES}} }}

\def\de        {\ensuremath {\Delta E}}

\def\avbr      {\ensuremath{{\BR}[B\rightarrow(\rho/\omega)\gamma]}}
\def\VtdVts    {\ensuremath{|V_{td}/V_{ts}|}}

\def\BFrp{1.10^{+0.37}_{-0.33}\pm 0.09} 
\def\BFrz{0.79^{+0.22}_{-0.20}\pm 0.06} 
\def\BFom{0.40^{+0.24}_{-0.20}\pm 0.05} 
\def\IST{-0.35 \pm 0.27} 
\def\BFav{1.25^{+0.25}_{-0.24}\pm 0.09} 
\def\BFavrhorho{1.36^{+0.29}_{-0.27}\pm0.10} 
\def\significance{6.4}                  
\def\VtdVtsval{0.200 ^{+0.021} _{-0.020} \pm 0.015 } 

\def\BFrpBelle{0.55^{+0.42+0.09}_{-0.36-0.08}} 
\def\BFrzBelle{1.25^{+0.37+0.07}_{-0.33-0.06}} 
\def\BFomBelle{0.56^{+0.34+0.05}_{-0.27-0.10}} 
\def\BFavBelle{1.32^{+0.34+0.10}_{-0.31-0.09}} 

\def\ifb{\;\mbox{fb}^{-1}}

\def\numBB{347 million}

\def\GeV{\;\mbox{GeV}}

\def\de   {\Delta E}

\def\bkg  {B \to K^{*}\gamma}

\def\brpg {B^+ \to \rho^+\gamma}
\def\brzg {B^0 \to \rho^0\gamma}
\def\bomg {B^0 \to \omega\gamma}

\begin{document}

\title{Radiative Penguin Decays}

%

\author{J\"urgen Kroseberg}
\affiliation{Santa Cruz Institute for Particle Physics, 
University of California,\newline
1156 High Street, Santa Cruz, CA 95064, USA}

\begin{abstract}
Selected recent results from experimental studies of radiative penguin decays of $B$ mesons 
by the Belle and \babar\ collaborations are discussed: preliminary findings from a first 
inclusive measurement of \mbox{$b\to s\gamma$} using a hadronic tag by \babar, first 
presented at this conference, updated preliminary results Belle  on $B\to p\bar{\Lambda}\gamma$, 
first shown at the Moriond QCD workshop earlier this year, and a recently published \babar\ study 
of $B\to\rho\gamma$ and $B\to\omega\gamma$ decays.
\end{abstract}

\maketitle

\thispagestyle{fancy}


\section{Introduction}
\begin{figure}[!h]
\begin{center}
\hspace*{-3.5mm}\includegraphics[width=200pt]{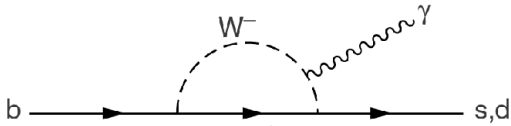}
\caption{Example leading-order Feynman diagram for $b\to s\g$ and $b\to d\g$ 
transitions.}
\label{fig:feyndiag} 
\end{center}
\end{figure}

Within the standard model of particle physics (SM), the rare, flavor-changing-neutral-current decays 
\mbox{$b\to d\gamma$} and \mbox{$b\to s\gamma$} are forbidden at tree level. The leading-order 
processes are one-loop electroweak penguin diagrams as shown in Figure~\ref{fig:feyndiag},  where 
the top quark is the dominant virtual quark contribution. In the context of theories beyond the SM, 
new virtual particles may appear in the loop, which could lead to measurable effects on experimental 
observables such as branching fractions and $CP$ asymmetries~\cite{bsm}.

The shape of the photon energy spectrum is insensitive to non-SM physics~\cite{Kagan:1998ym} 
but can be used to determine the Heavy Quark Expansion (HQE) parameters \mb and \mupisq,
which are related to the mass and momentum of the \b quark within the $B$ meson. Improved 
measurements of these parameters are relevant to, e.g., reduce the error in the CKM matrix 
elements $\Vcb$ and $\Vub$ determined from semi-leptonic \B-meson decays~\cite{Buchmuller:2005zv}.

In the following, three recent experimental studies of radiative penguin decays are discussed 

\section{Measuring $\boldmath{b\to s\gamma}$  with a hadronic tag}
Previous inclusive measurements of $b\to s\gamma$ transitions, by the CLEO~\cite{Chen:2001fj},
Belle~\cite{Koppenburg:2004fz} and \babar\ ~\cite{BABARSEMI,BABARINCL} collaborations, used 
either fully inclusive event samples, requiring only a high-energy lepton in the event or 
no tag at all, or a combination of several exclusive decay modes. A new (preliminary) analysis, 
based on \babar\ data corresponding to an integrated luminosity of $210\ifb$,  uses for the first 
time a recoil method to select candidate $B$ decays, where \BB events are tagged by a fully 
reconstructed hadronic decay of one the $B$ mesons, in the following referred to as the 
{\it tag} $B$, and radiatively decaying {\it signal} $B$ mesons are reconstructed from the 
remaining particles in the event. While the hadron tag efficiency is low (about $0.3\%$), this 
method allows for the inclusive study of $b\to s\gamma$ decays in a relatively clean environment and 
a determination of the momentum, charge, and flavor of the \B mesons. Thus, it is possible to 
measure the photon spectrum in the rest frame of the signal \B, to separate charged and neutral 
\B mesons and to determine the CP-asymmetry \acp.
\begin{figure*}[!ht]
\begin{center}
\includegraphics[height=2.4in]{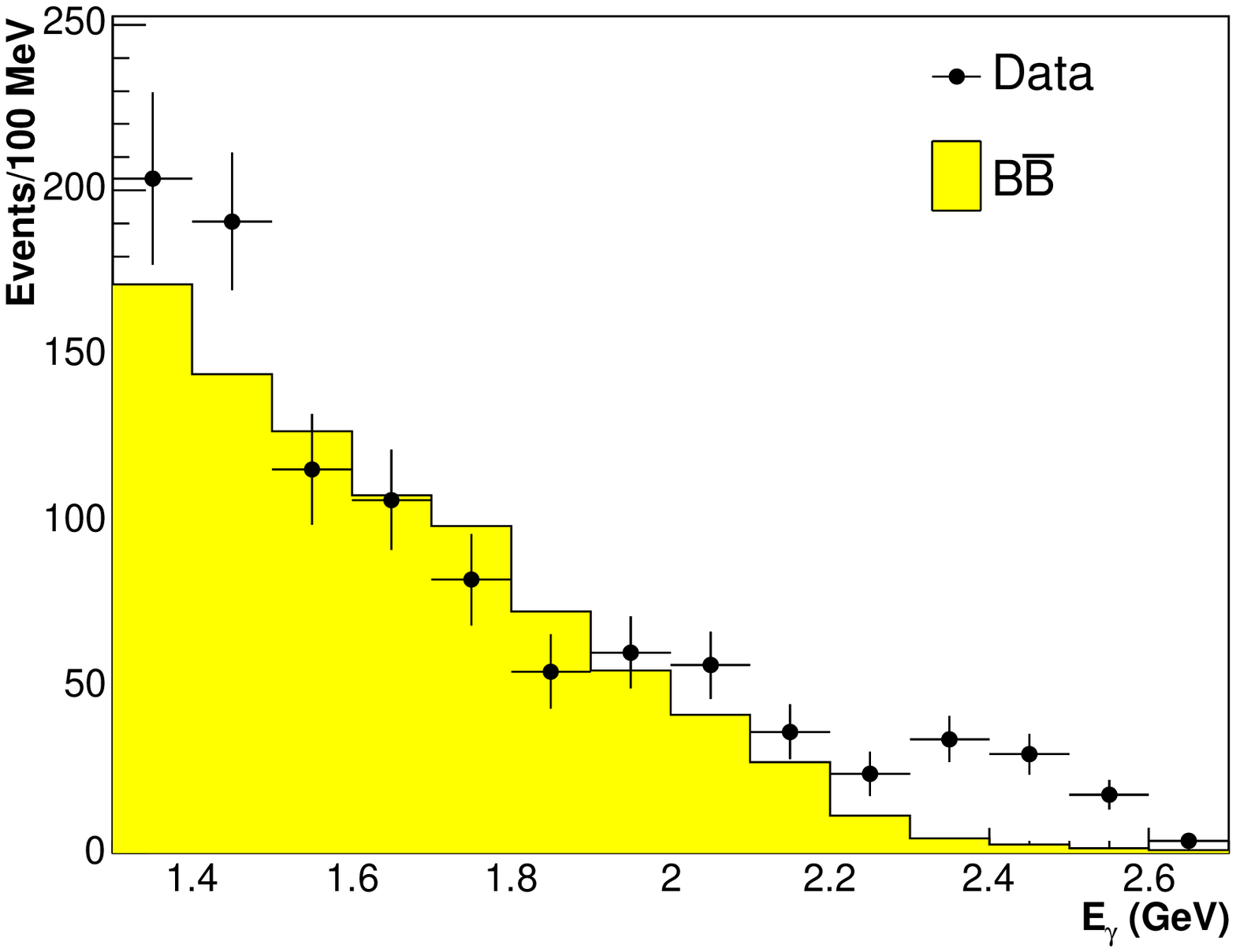}\hfill
\includegraphics[height=2.4in]{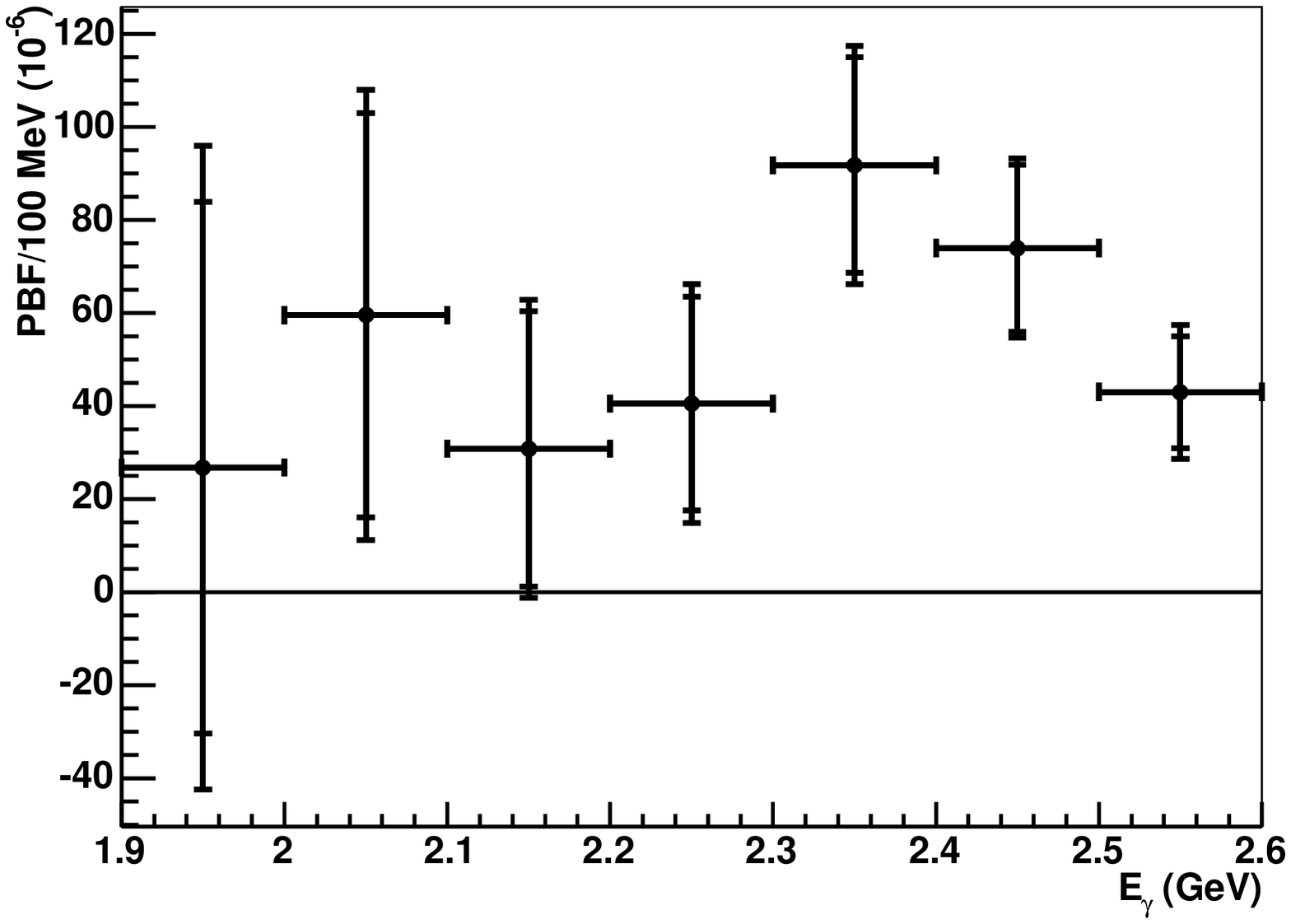}
\end{center}

\setlength{\unitlength}{1cm}    
\begin{picture}(0.1,0.1)
\put(-5.5,4.9){\includegraphics[height=0.26in]{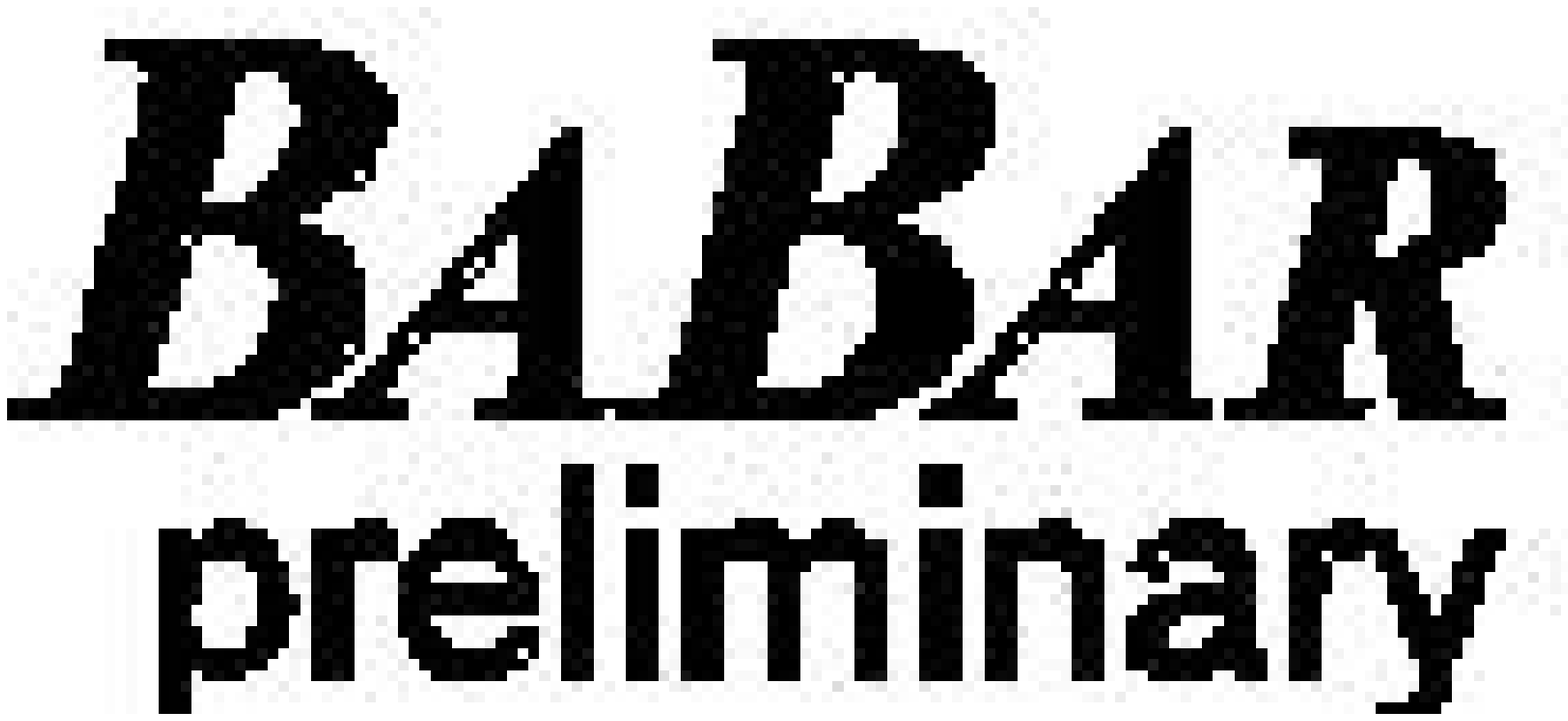}}
\put(4.8,1.7){\includegraphics[height=0.26in]{bbprelim.eps}}
\end{picture}
\caption{Left: the measured numbers of $B$ events as a function of photon
energy.  The points are from data; the histogram is from a \BB Monte Carlo
sample which excludes the signal decay \bxsg; Right: the partial branching fractions 
$(1/\Gamma_B) (d\Gamma/dE_{\gamma})$  with statistical and total error. 
All results are preliminary}
\label{fig:sgspect}
\end{figure*}

In order to reconstruct a large sample of tag $B$ mesons, hadronic decays
to $\Db Y^{\pm}$ and $\Db^* Y^{\pm}$ final states are selected, where $Y^{\pm}$ 
denotes relevant combinations of mesons -- $\pi^{\pm}$, $K^{\pm}$, $\KS$, 
and $\piz$ -- with a total charge of $\pm 1$~\cite{Aubert:2003zw}. 
Those particles in the event that are not 
reconstructed as part of the tag $B$ are required to include an isolated, well-reconstructed 
photon candidate with an energy $E_{\gamma} > 1.3 \gev$ in the signal $B$ meson rest frame. 
Photon candidates that are found to be consistent with stemming from \piz or $\eta$ decays 
are rejected. Continuum backgrounds ($\ep\en \to q\bar{q}$, with $q=u,d,s,c$) are suppressed 
using a Fisher discriminant combining twelve quantities that are sensitive to event 
shape differences between $B$ decays and continuum processes. 

\begin{table*}[!ht]
\begin{center}
\caption[]{\label{tab:sgresults} Preliminary results for the $b\to s\gamma$ branching fraction and 
moments of the photon energy spectrum, with statistical and systematic errors, for different 
minimum photon energies \ecut.\\ \ } 
\begin{tabular}{ccccccccccccc}
\hline
\ecut (\gev) &  \ &
\multicolumn{3}{c}{\vvsmallrule $\mathcal{B}(\bxsg)$  $(10^{-6})$}& \ &
\multicolumn{3}{c}{\vvsmallrule \meg  (\gev)}   & \ &
\multicolumn{3}{c}{\vvsmallrule \vegs  ($\gev^2$)}\\\hline

1.9 &                                          \hspace{4em} &	
366 &	$\pm$	85  &   $\pm$   59   &         \hspace{4em} &
      2.289 &	$\pm$  0.058&	$\pm$  0.026 & \hspace{4em} &
      0.0334&	$\pm$  	0.0124& $\pm$   0.0263 \\
2.0 &   \ &
	339 &	$\pm$	64  &	$\pm$   47 & \ &
  	2.315&	$\pm$  0.036&	$\pm$  0.020& \ &
 	0.0265&	$\pm$  	0.0057& $\pm$   0.0203\\
2.1 &   \ &
	278 &	$\pm$	48  &	$\pm$   34 & \ &
       2.371&	$\pm$  0.025&	$\pm$  0.011& \ &
	0.0142&	$\pm$  	0.0037& $\pm$   0.0111\\
2.2 &  \ &
	248 &	$\pm$	38  &	$\pm$	26 & \ &
	2.398&	$\pm$  0.016&	$\pm$  0.006& \ &
	0.0092&	$\pm$  	0.0015& $\pm$   0.0061\\
2.3 &  \ &
	207 &	$\pm$	30  &	$\pm$	19 & \ &
	2.427&	$\pm$  0.010&	$\pm$  0.007& \ &
0.0059&	$\pm$  	0.0007& $\pm$   0.0073\\\hline
\end{tabular}
\end{center}
\end{table*}

The numbers of remaining $B$ and non-$B$ events are determined by means of fits to the
beam-energy-substituted mass $\mes = \sqrt{s/4 - \vec{p}^{\,2}_B}$.\footnote{Here, 
$\sqrt{s}$ is the total energy in the center of mass (CM) frame, and $\vec{p}_B$ denotes 
the \B candidate CM momentum.}
The left hand side of Figure~\ref{fig:sgspect} shows the preliminary results as a function of 
the photon energy. The points are from the data; the solid histogram was obtained from a \BB 
Monte Carlo sample (excluding the signal decay \bxsg) and then 
scaled according to the results of a fit to the data in the region 
$1.3 \, \gev < E_{\gamma}< 1.9 \, \gev$.  
For $\eg > 1.9 \gev$ $119 \pm 22$ \bxsg signal events are observed over
a \BB background of $145 \pm 9$ events.

The differential decay rate $(1/\Gamma_B)(d\Gamma/dE_{\gamma})$ is measured in bins of the 
signal \B rest frame photon energy in the range $1.9 \le E_{\gamma} < 2.6$ GeV. For a given 
bin $i$ this is determined according to
\begin{equation}
\label{eq:estimator} 
\frac{1}{\Gamma_B} 
\frac{d\Gamma_i} {dE_{\gamma}}
= \frac{N_i - b_i} {\varepsilon_i N_B C_{\rm tag}} \;,
\end{equation}
where $N_i$ is the number of $B$ events in the bin, $b_i$ is the number of $B$ mesons from decays 
other than $b \rightarrow (s,d)\gamma$, and $N_B$ is the total number of $B$ mesons in the sample. 
The efficiency $\varepsilon_i$ corrects for both acceptance and bin-to-bin resolution effects, and 
the correction factor $C_{\rm tag}$ accounts for any dependence of the hadronic \B tag probability 
on the the presence of a \bxsg final state.

In the right part of Figure~\ref{fig:sgspect} the partial branching fraction
is shown after all corrections. The systematic uncertainties -- mainly arising from 
the modelling of the \BB background, the \mes fit parametrization, the description of 
the detector response, and the dependence on the \bxsg signal model -- are included here.
The results for the integrated branching fraction and moments of the photon energy
spectrum above different minimum photon energies \ecut\, are summarized in 
Table~\ref{tab:sgresults}. 
Figure~\ref{fig:moments} shows the first and second moments of the
photon energy spectrum as a function of $E_{\rm cut}$; good agreement with 
previous measurements is found.

\begin{figure*}[!ht]
\begin{center}
  \mbox{\includegraphics[width=3.0in]{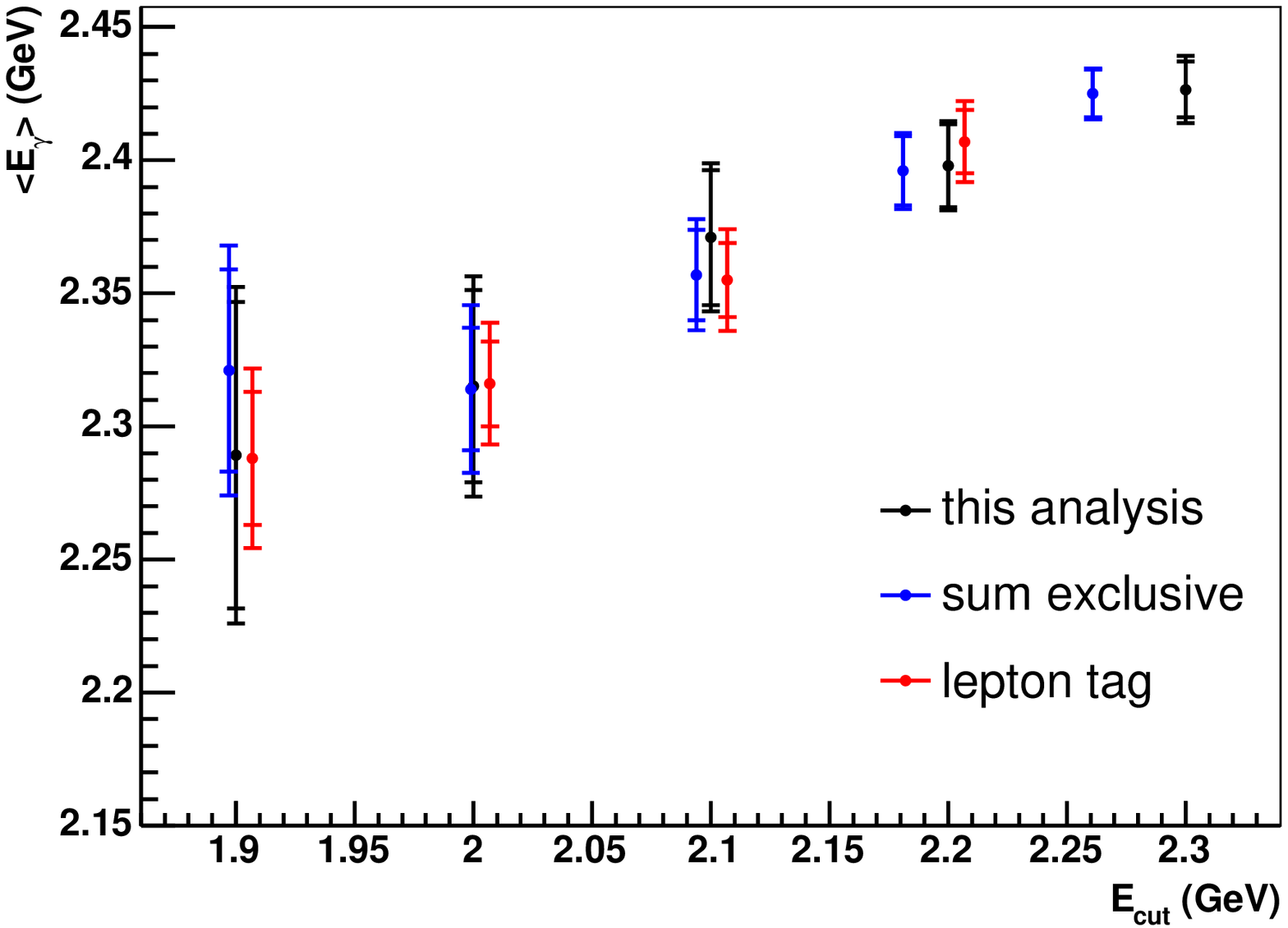}}
  \mbox{\includegraphics[width=3.0in]{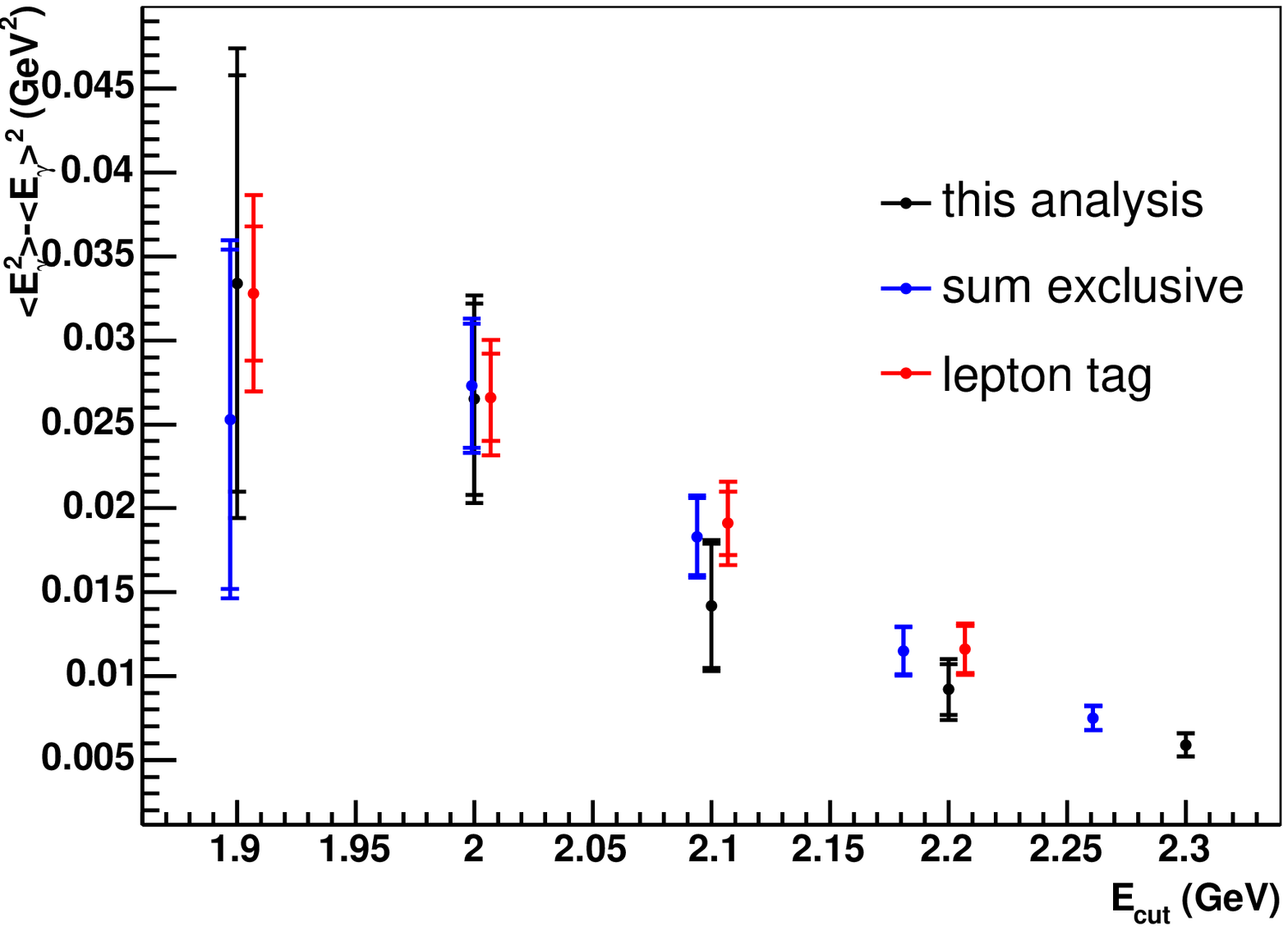}}
\end{center}

\setlength{\unitlength}{1cm}    
\begin{picture}(0.1,0.1)
\put(-5.5,1.9){\includegraphics[height=0.26in]{bbprelim.eps}}
\put(2.0,4.7){\includegraphics[height=0.26in]{bbprelim.eps}}
\end{picture}

\vspace*{-0.5cm}
\caption{First and second moments of the photon energy spectrum as a 
function of $E_{\rm cut}$.  For comparison, the results from~\cite{BABARSEMI} 
and~\cite{BABARINCL} are show as well.}
\label{fig:moments}
\end{figure*}

Using  \cite{Buchmuller:2005zv} to extrapolate the measured branching fraction down 
to a minimum photon energy of $1.6\GeV$ yields 
\begin{equation}
{\cal B}(b\to s\gamma)|_{E_\gamma>1.6\gev}=(3.9\pm0.9\pm0.6)\times 10^{-4}.
\end{equation}
This is in good agreement with the current experimental world average~\cite{hfag2006} of 
$(3.55 \pm 0.26) \times 10^{-4}$ as well as recent NNLO QCD 
calculations~\cite{Misiak:2006zs,Becher:2006qw,Andersen:2006hr}.

All these preliminary results are currently limited by statistics; adding more data 
in future measurements is expected to significantly reduce also the systematic uncertainties.  
It should also be noted that the recoil method is complementary to those of other 
measurements of $b\to s\g$ transitions; the largely independent systematic uncertainties 
will facilitate a combination of the results.

\section{Study of  $B\to p\bar{\Lambda}\gamma$ decays}
The decay $B^+\to p\bar{\Lambda}\gamma$ 
was first observed by Belle in 2005 based on a 
data sample corresponding to an integrated luminosity of $140\ifb$~\cite{oldlambda}. 
These results have now been updated with a preliminary measurement\footnote{
Since this conference, this analysis has been finalized and published~\cite{newlambda}.}
 using $414\ifb$ of 
Belle data.  This is a rare process; the  SM branching fraction is predicted to be 
$\approx1\times10^{-6}$. The final state baryons restrict 
the phase space, thus giving access to low photon momenta. The experimental 
study of this decay provides information on the baryon production mechanism. 
For example, the spin of the $s$ quark can be probed through a helicity analysis; large 
wrong-helicity contributions would indicate physics beyond the SM.

Figure~\ref{fig:lambdasig} shows the projections of a two-dimensional fit to the distributions 
of the energy difference $\de \equiv E^*_{B}-E_{\rm beam}^*$, where $E^*_B$ is the CM energy of 
the $B$ meson candidate and $E_{\rm beam}^*$ is the CM beam energy, and \mes for selected events 
with an invariant di-baryon mass $m_{p\bar{\Lambda}}<2.8\gev$. 
\begin{figure}[!b]
\begin{center}
\hspace*{-3.5mm}\includegraphics[width=230pt]{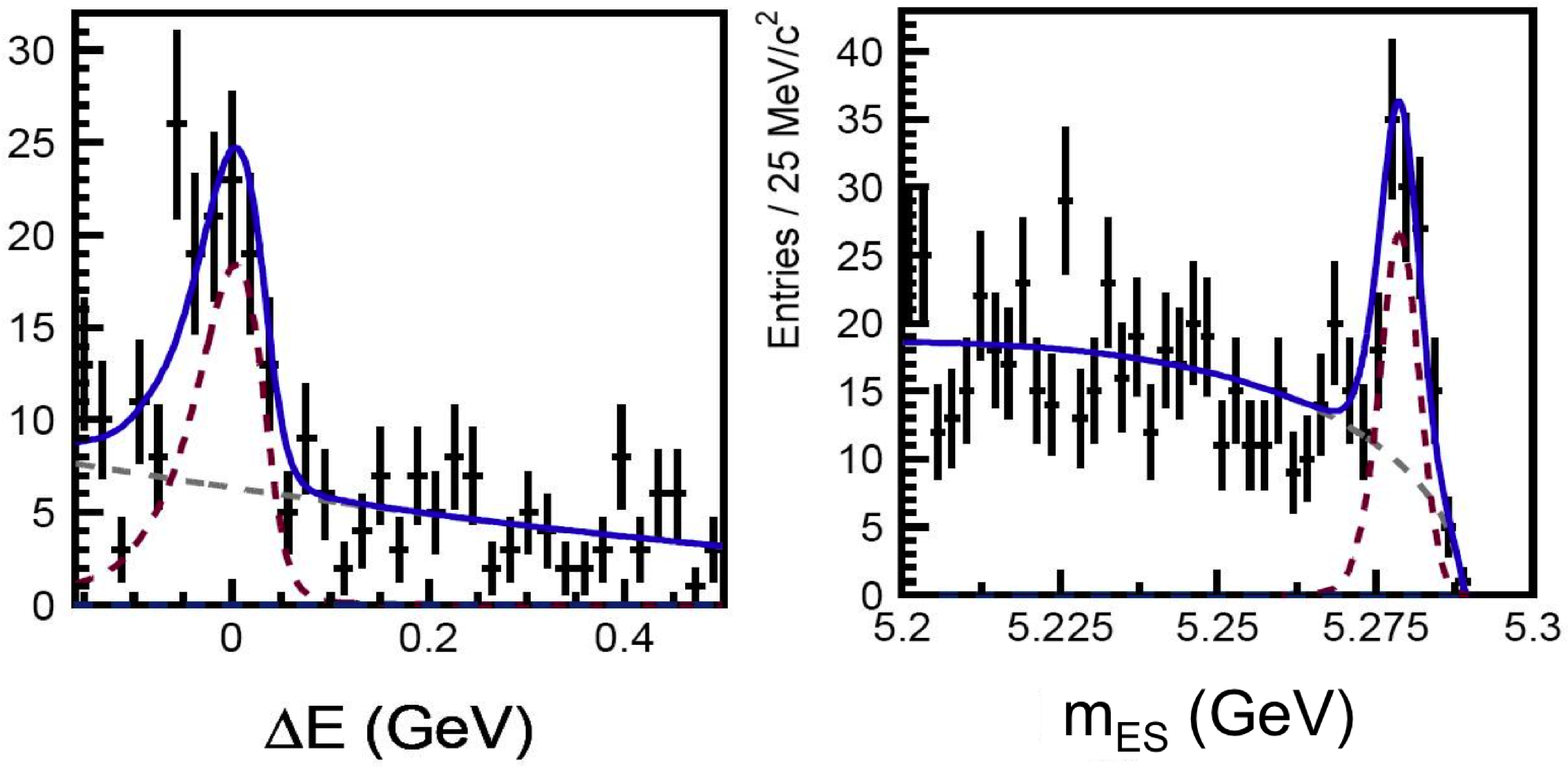}
\setlength{\unitlength}{1cm}    
\begin{picture}(0.1,0.1)
\put(-6.5,2.9){\includegraphics[height=0.30in]{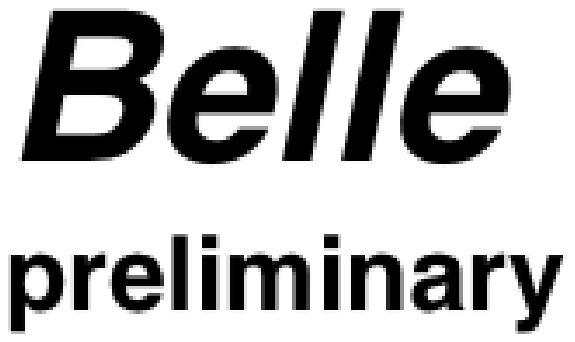}}
\put(-2.5,2.9){\includegraphics[height=0.30in]{blprelim.eps}}
\end{picture}
\caption{The $\Delta E$ (left) and \mes (right) distributions for selected 
$B^+\to p\bar{\Lambda}\gamma$ in the $p\bar{\Lambda}$ invariant mass region $<2.8\gev$, 
together with the result of the fit (solid curve), which is also shown separated into 
signal (dash-dotted) and background (dashed) components.
\label{fig:lambdasig} }
\end{center}
\end{figure}
About 100 signal events are found, 
corresponding to a statistical significance of $14\sigma$. Turning this into a branching fration and
extrapolating to the full di-baryon mass range yields 
\begin{equation}
{\cal B}=(2.45^{+0.44}_{-0.38}\pm{0.22})\times 10^{-6}, 
\end{equation}
somewhat above current SM predictions~\cite{Cheng,Geng}.

A near-threshold enhancement is found in the di-baryon mass distribution, see the left part of 
Figure~\ref{fig:lambdadist}.
\begin{figure}[!h]
\begin{center}
\hspace*{-0.3cm}\includegraphics[height=109pt]{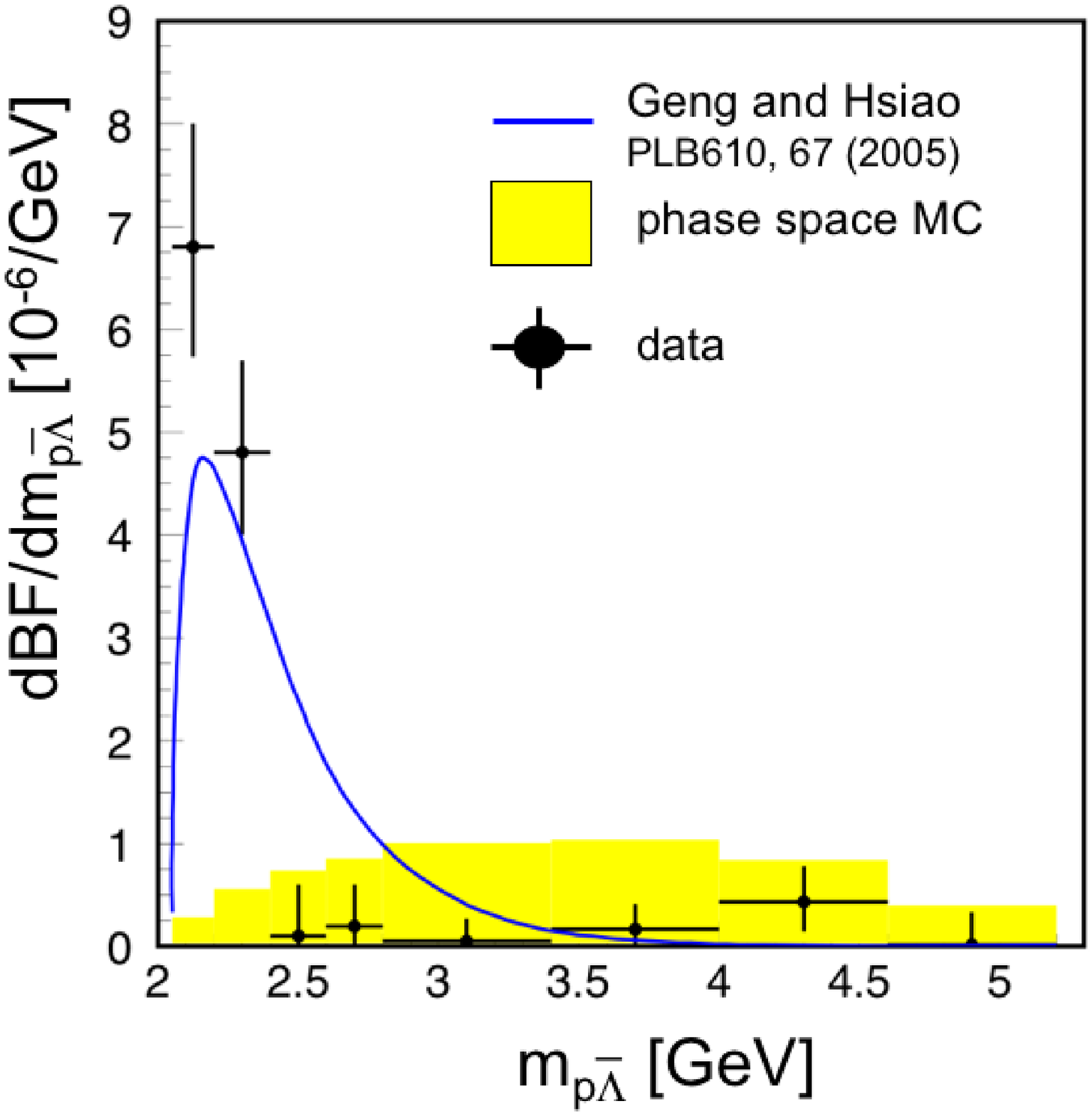}
\hspace*{-0.0cm}\raisebox{0.08cm}{\includegraphics[height=110pt]{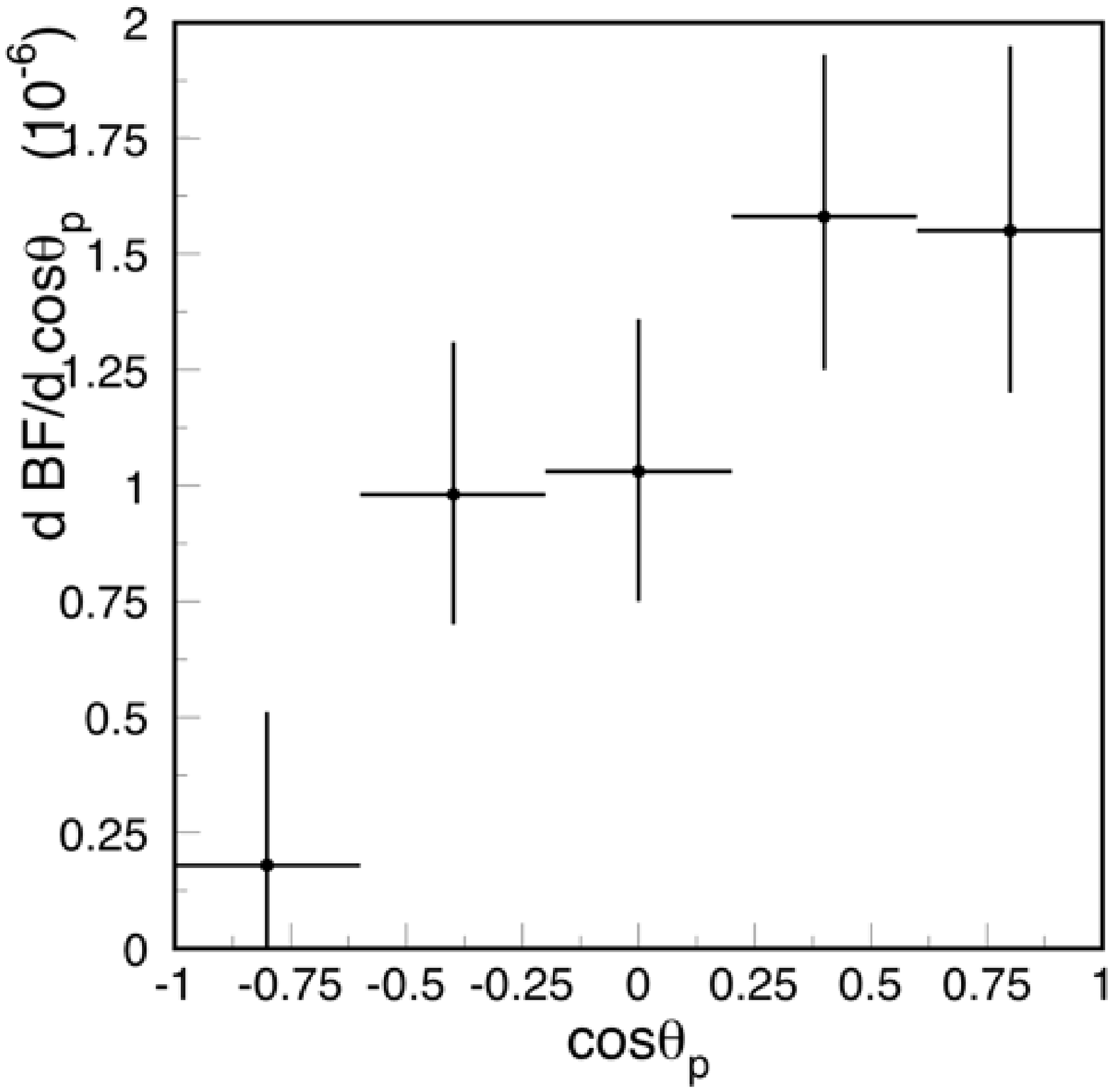}}
\setlength{\unitlength}{1cm}    
\begin{picture}(0.1,0.1)
\put(-1.8,1.5){\includegraphics[height=0.30in]{blprelim.eps}}
\put(2.3,1.5){\includegraphics[height=0.30in]{blprelim.eps}}
\end{picture}
\caption{Measured distributions of the $p\bar{\Lambda}$ invariant mass (left), compared to 
theoretical expectations, and $\cos(\theta_p)$, where the helicity angle $\theta_p$ is the 
angle between proton and photon direction in the di-baryon rest frame.   
\label{fig:lambdadist}} 
\end{center}
\end{figure}
This was previously observed for other baryonic $B$ decays and has since then been the focus 
of considerable theoretical interest. 

The right hand side of Figure~\ref{fig:lambdadist} shows the measured distribution of 
$\cos(\theta_p)$, where the helicity angle $\theta_p$ is the angle between proton and photon 
direction in the di-baryon rest frame.   It is consistent with a short-distance $b\to s\gamma$ 
description of this decay. The asymmetry with respect 
to $\cos(\theta_p)=0$ is determined to be $A_\theta=0.29\pm 0.14$. The CP asymmetry, 
$A_{CP}=0.17\pm 0.17$, is found to be consistent with zero.

Table~\ref{tab:lpgresults} summarizes these results and includes, for comparison, corresponding 
measurements for the decays $B^{+}\to p\bar{\Lambda}\pi^0$, which is observed for the first 
time,  and $B^{0}\to p\bar{\Lambda}\pi^-$.

\begin{table}[!t]
\centering
\caption{Measured branching fractions ${\cal B}$, angular asymmetries $A_\theta$, 
and CP asymmetries $A_{CP}$. All results are preliminary.\label{tab:lpgresults} } 
\renewcommand{\arraystretch}{1.3}
\begin{tabular}{l|c@{\hspace{0.2cm}}c@{\hspace{0.2cm}}c}
\hline
\ \hfill Mode\hfill\  & ${\cal B}$ $(10^{-6})$ & $A_\theta$ & $A_{CP}$  \\\hline
$B^{+}\to p\bar{\Lambda}\gamma$ & $2.45^{+0.44}_{-0.38}\pm{0.22}$ &$+0.29\pm 0.14$ &$+0.17\pm 0.17$\\
$B^{+}\to p\bar{\Lambda}\pi^0$  & $3.00^{+0.61}_{-0.53}\pm{0.33}$ &$-0.16\pm 0.18$ &$+0.01\pm 0.17$\\
$B^{0}\to p\bar{\Lambda}\pi^-$  & $3.23^{+0.33}_{-0.29}\pm{0.29}$ &$-0.41\pm 0.11$ &$-0.02\pm 0.10$\\ 
\hline
\end{tabular}
\end{table}

\section{$B\to\rho/\omega\gamma$ branching fractions}
With respect to \mbox{$b\to s\gamma$}, the corresponding \mbox{$b\to d\gamma$} branching fractions 
are suppressed by a factor $\approx0.04$; the SM branching 
fractions for the decays of $B$ mesons to $\rho\gamma$ and $\omega\gamma$ are ${\cal O}$($10^{-6}$).
While the calculations of the exclusive decay rates have large uncertainties due to non-perturbative
long-distance QCD effects, some of this uncertainty cancels in the ratio of $B$ \to $\rho/\omega$\g 
to \mbox{$B$ \to $K^{*}$\g} branching fractions. Since the dominant diagram involves a virtual top 
quark, this ratio is related to the ratio of Cabibbo-Kobayashi-Maskawa (CKM) matrix elements
$\VtdVts$~\cite{alivtdvtstheory,ali2004,Bosch:2004nd} via\footnote{
The coefficient $\zeta$ is the ratio of the form factors for the decays $B \rightarrow \rho\gamma$
and $B \rightarrow K^{*}\gamma$ and $\Delta R$ accounts for different dynamics in the decay.
}
\begin{equation}\label{eq:vtd}
\frac{\avbr}
{{\cal B}(B \rightarrow K^{*}\gamma)}=
\left| \frac{V_{td}}{V_{ts}} \right|^{2}
\left(\frac{1-m_{\rho}^{2}/M_{B}^{2}}{1-m_{K^{*}}^{2}/M_{B}^{2}}\right)^{3}
\zeta^{2} [1+\Delta R].
\end{equation}
Physics beyond the Standard Model could affect these decays, creating inconsistencies between 
this measurement of $\VtdVts$ and that obtained from the ratio of $B^0$ and $B^0_s$ mixing 
frequencies~\cite{bsmixing}.

After previous searches by $\babar$~\cite{oldbabar} and CLEO \cite{cleo}, which 
found no evidence for the decays {\ensuremath{B\rightarrow\rho\gamma}} and 
{\ensuremath{B\rightarrow\omega\gamma}}, an observation of the decay \mbox{$\brzg$} 
was reported by the Belle collaboration~\cite{newbelle}.
A recently published \babar\ study~\cite{newbabar} of the decays $\Bp\to\rhop\gamma$, 
$\Bz\to\rhoz\gamma$, and $\Bz\to\omega\gamma$ is summarized in the following.

From a data sample containing \numBB\ \BB\ pairs, which corresponds to an integrated luminosity 
of \mbox{316 fb$^{-1}$}, the decays {\ensuremath{B\rightarrow\rho\gamma}} and 
{\ensuremath{B\rightarrow\omega\gamma}}  are reconstructed by combining a high-energy photon 
with a vector meson reconstructed in the decay modes $\rho^0\to\pip\pim$, $\rho^+\to\pip\piz$,  
and $\omega\to\pip\pim\piz$. The dominant source of background is continuum events
($\ep\en \to q\bar{q}$, with $q=u,d,s,c$) that contain a high-energy photon from $\piz$ or 
$\eta$ decays. Other backgrounds include photons from initial-state radiation processes,
decays of $\bkg$ ($\Kstar\rightarrow K\pi$), decays of $\B\rightarrow(\rho/\omega)\piz$ or 
$\B\rightarrow(\rho/\omega)\eta$ and combinatorial background  from higher-multiplicity 
$\b\rightarrow\s\gamma$ decays.

High-energy photons $\gamma$ for which the combination with another photon $\gamma'$ is found -- 
based on a likelihood ratio constructed from  $E_{\gamma'}$ and  $m_{\gamma\gamma'}$ -- to be 
consistent with a $\piz$ or $\eta$ decay  are rejected. A neural network, exploiting differences 
between background and \B decays in, e.g., lepton and kaon production (through flavor-tagging 
variables described in ~\cite{babartag}) and event shape, is used to suppress continuum background 
events.
\begin{figure}
\includegraphics[width=0.5\linewidth,clip=true]{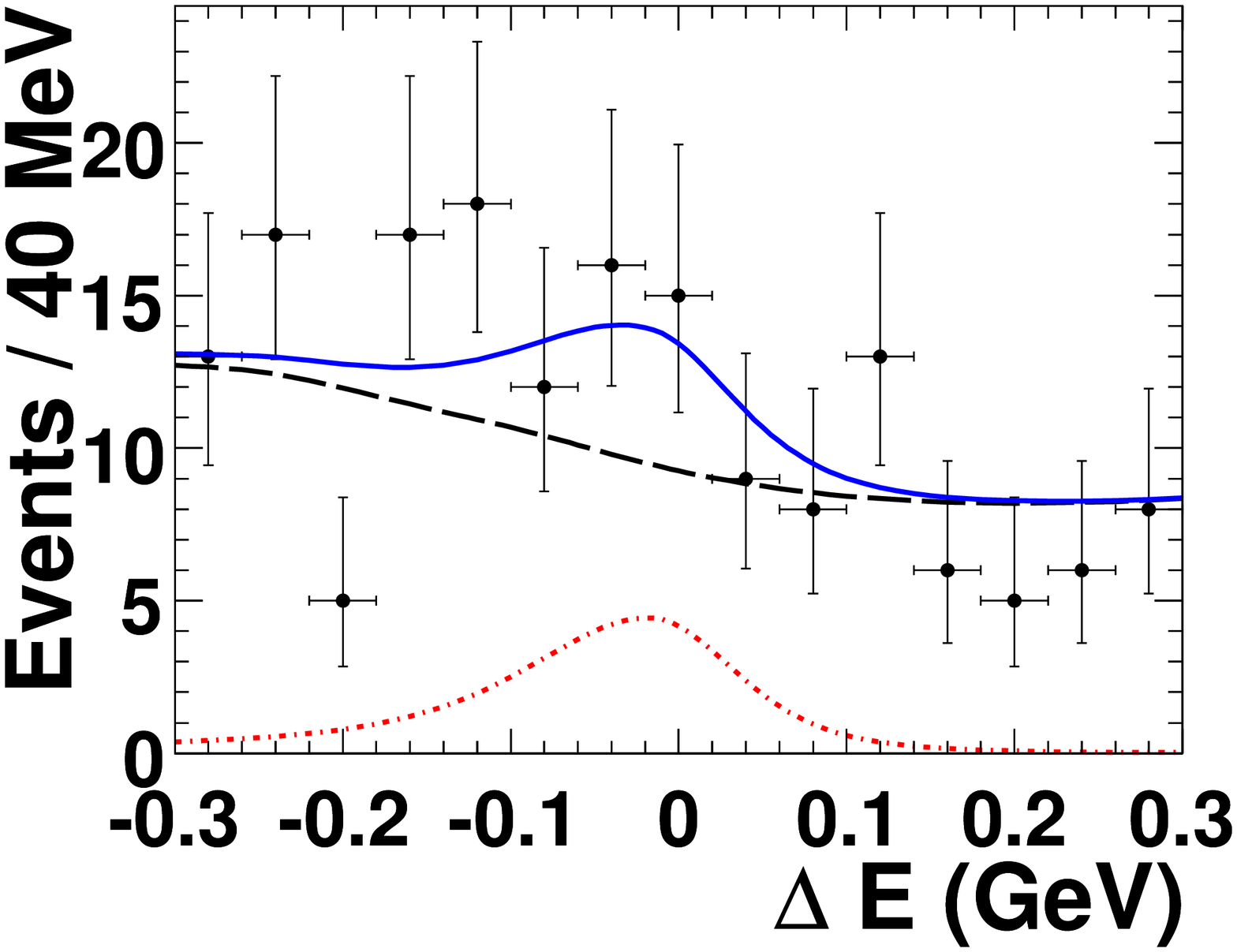}%
\includegraphics[width=0.5\linewidth,clip=true]{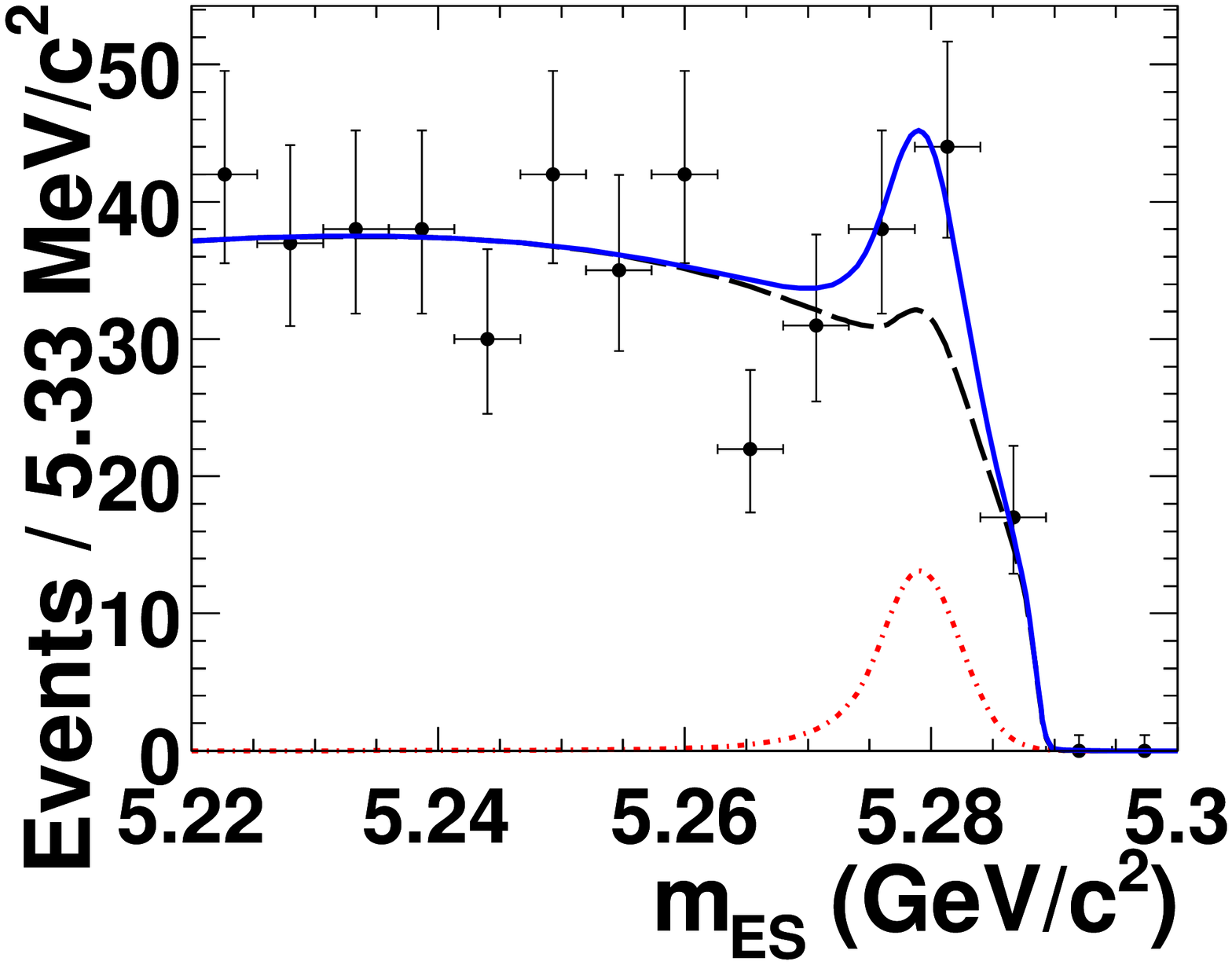}\\
\includegraphics[width=0.5\linewidth,clip=true]{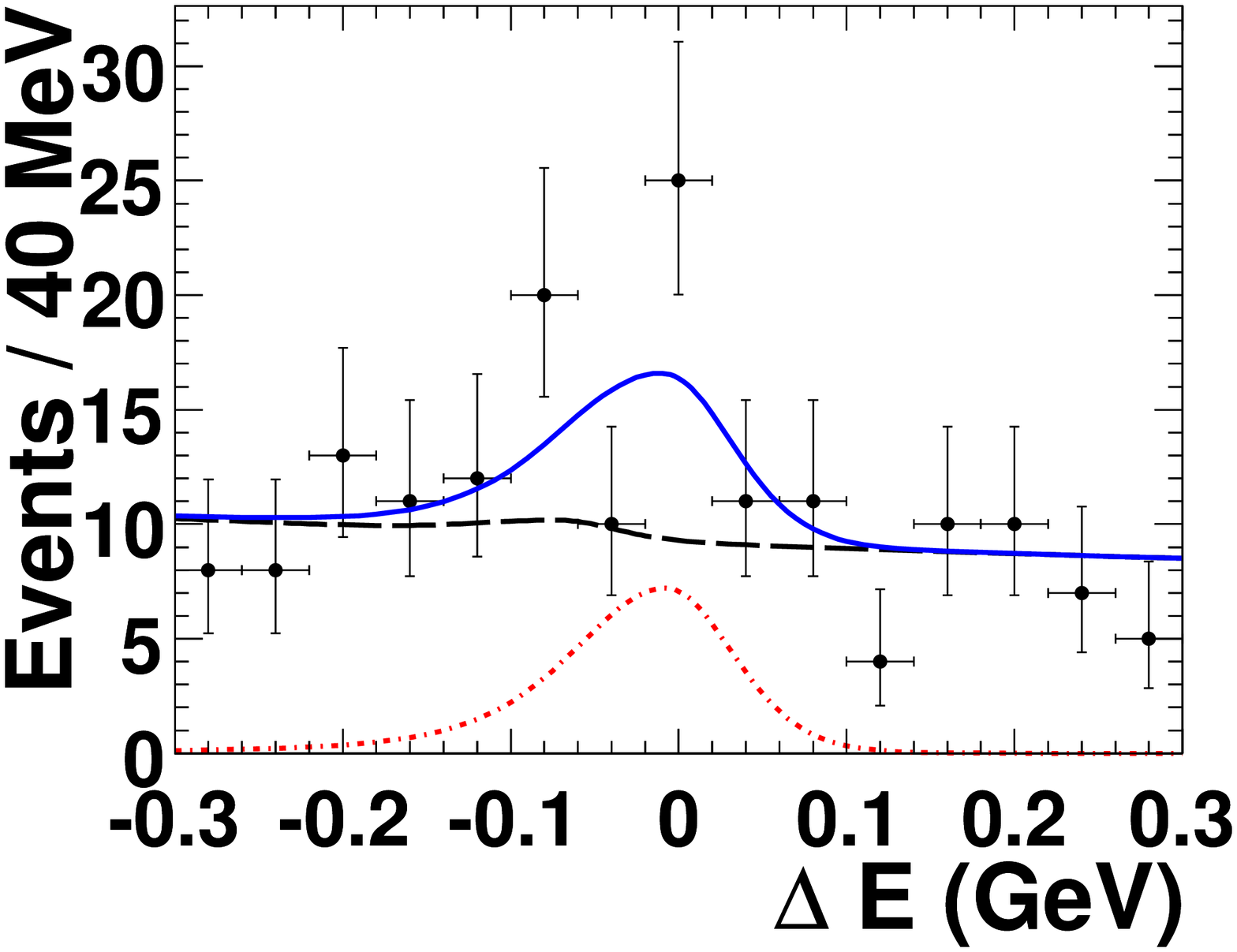}%
\includegraphics[width=0.5\linewidth,clip=true]{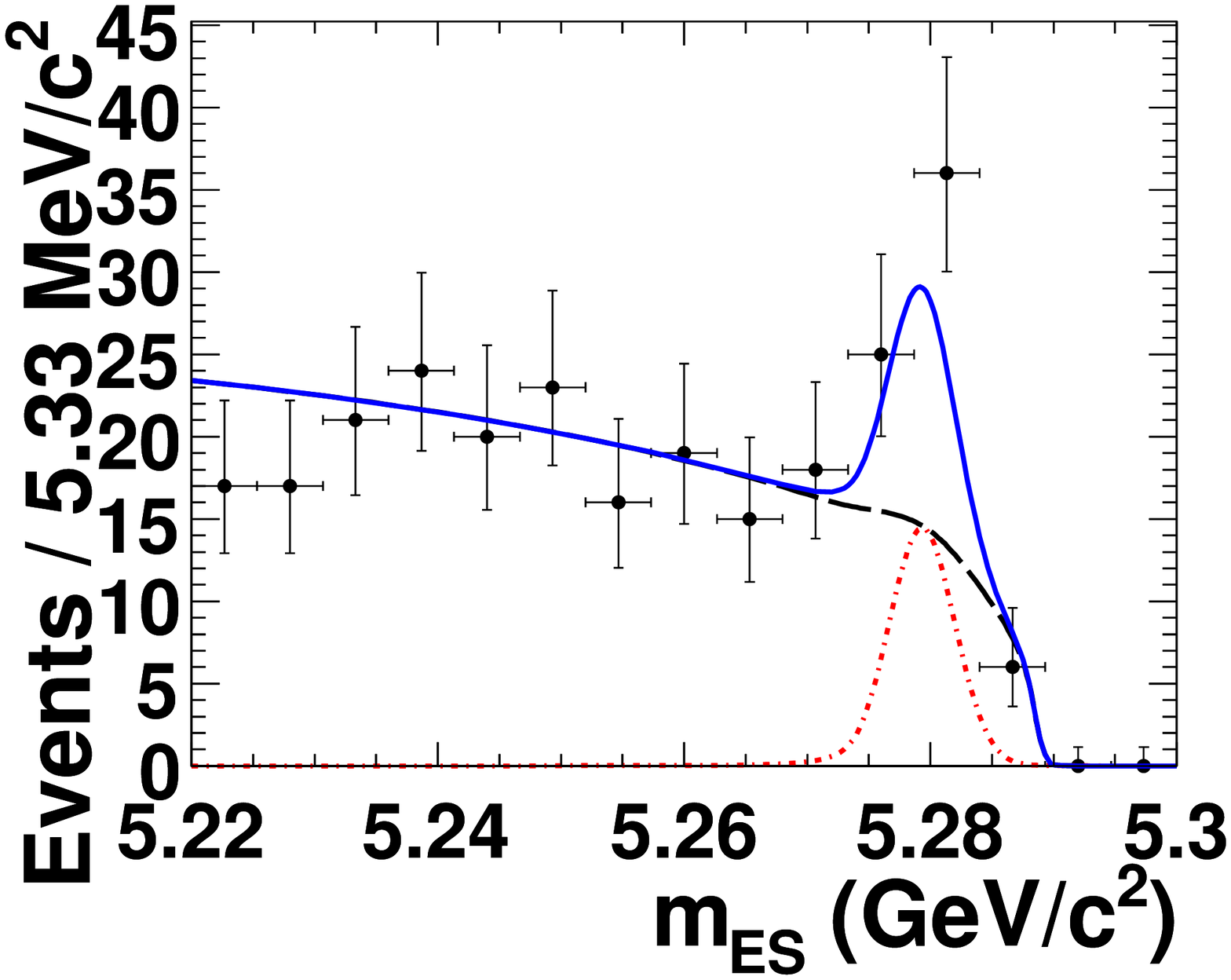}\\
\includegraphics[width=0.5\linewidth,clip=true]{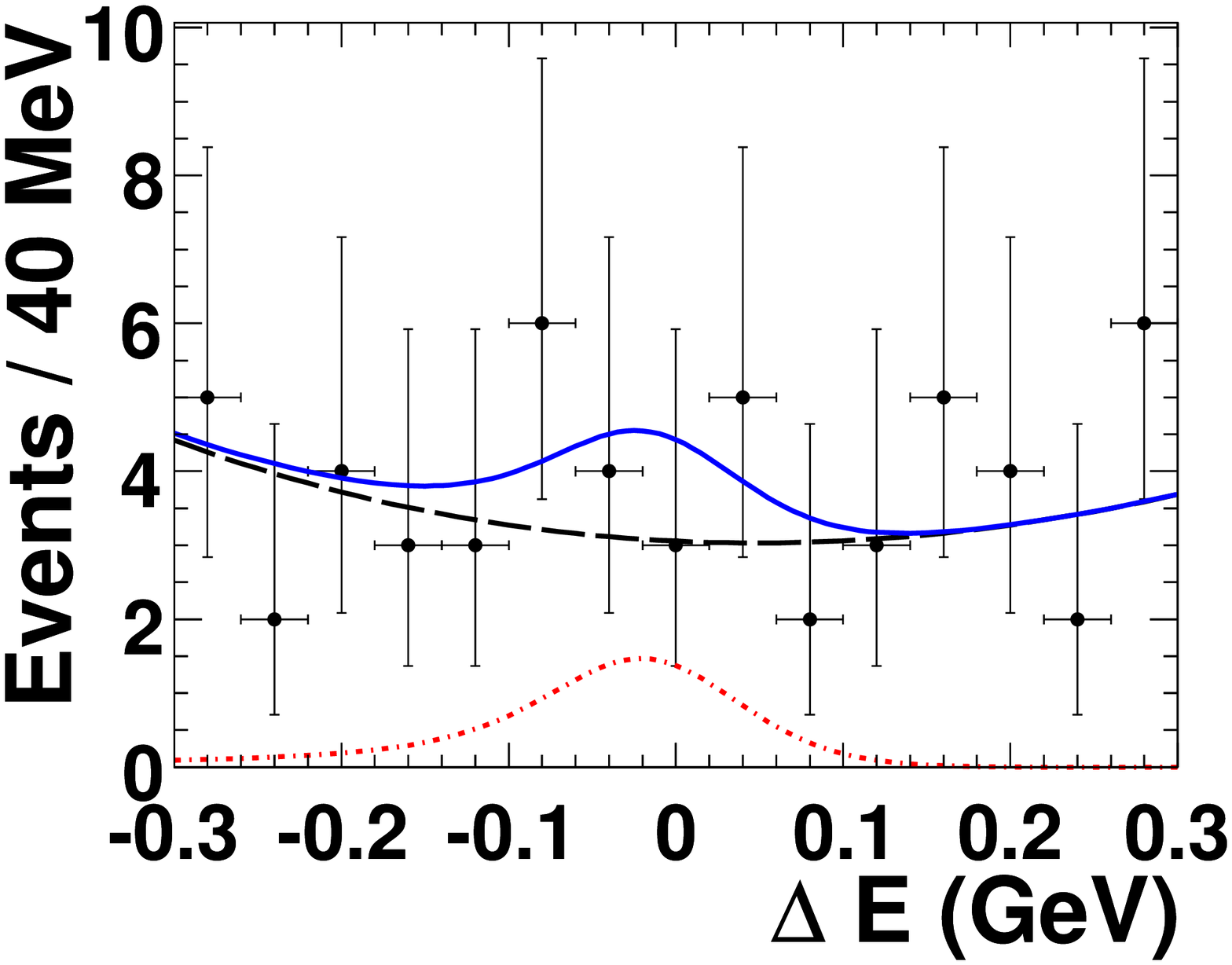}%
\includegraphics[width=0.5\linewidth,clip=true]{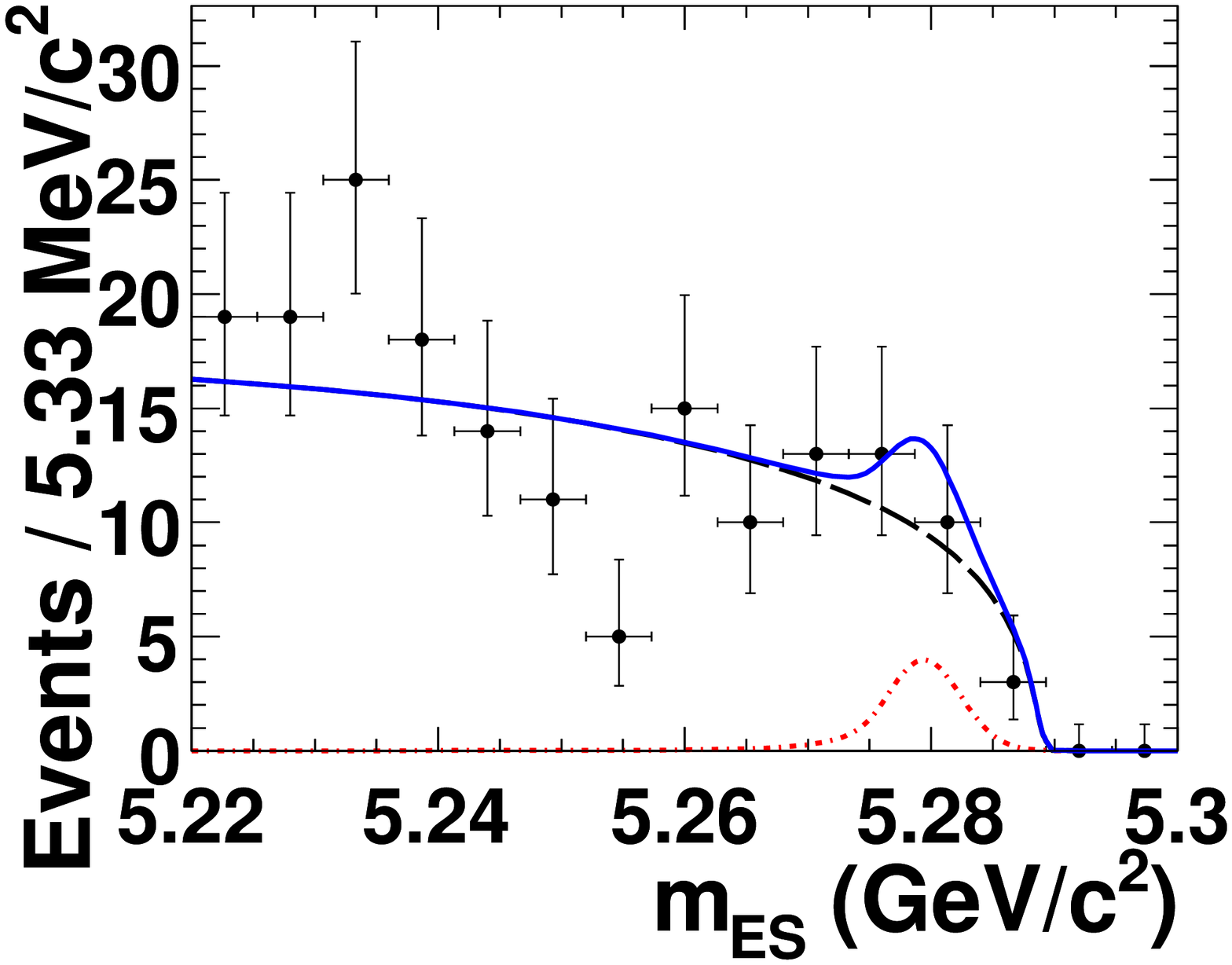}\\
\setlength{\unitlength}{1cm}    
\begin{picture}(0.1,0.1)
\put(-1.5,9.7){\includegraphics[height=0.12in]{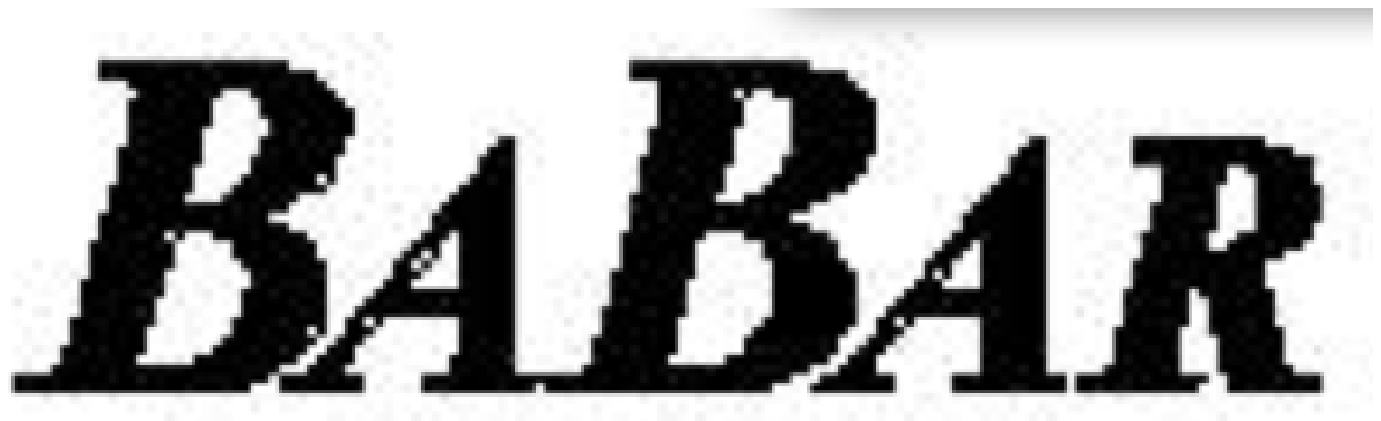}}
\end{picture}
\caption{
$\de$ and $\mes$ projections of the fits for the decay modes $\brpg$ (top),
$\brzg$ (middle), and $\bomg$ (bottom).
In each plot, the signal fraction is enhanced by selections  on the other fit variables.
The points are data, the solid line is the total of all contributions, and
the long-dashed (dashed-dotted) line is background-only (signal-only).
}\label{fig:rhoChfit}
\end{figure}
\begin{table*}
\centering
\caption{\label{tab:rgresults} The signal yield $(n_{\mathrm{sig}})$,
significance ($\Sigma$) in standard deviations including systematic errors, efficiency $(\epsilon)$,
and branching fraction $(\mathcal{B})$ for each mode.
The errors on $n_{\mathrm{sig}}$ are statistical only, while for the branching fraction
the first error is statistical and the second systematic.\\ \ }
\renewcommand{\arraystretch}{1.3}
\begin{tabular}{l|c@{\hspace{0.4cm}}c@{\hspace{0.4cm}}c|c@{\hspace{0.4cm}}c@{\hspace{0.4cm}}cc}
\cline{1-7}
&\multicolumn{3}{c|}{\babar}&\multicolumn{3}{c}{Belle} & \\
Mode & \hspace*{0.5cm}$n_{\mathrm{sig}}$ & $\Sigma$& $\mathcal{\B} (10^{-6})$\hspace*{0.5cm}  & \hspace*{0.5cm}$n_{\mathrm{sig}}$ & $\Sigma$& $\mathcal{\B} (10^{-6})$  &\\
\cline{1-7}
$\brpg$  & \hspace*{0.5cm}$42.0^{+14.0}_{-12.7}$     & $3.8\sigma$ & $\BFrp$\hspace*{0.5cm} & \hspace*{0.5cm}8.5 & $1.6\sigma$ &$\BFrpBelle$&\\
$\brzg$  & \hspace*{0.5cm}$38.7^{+10.6}_{-9.8}$      & $4.9\sigma$ & $\BFrz$\hspace*{0.5cm} & \hspace*{0.5cm}20.7& $5.2\sigma$ &$\BFrzBelle$&\\
$\bomg$  & \hspace*{0.5cm}$11.0^{+6.7}_{-5.6}$       & $2.2\sigma$ & $\BFom$\hspace*{0.5cm} & \hspace*{0.5cm}5.7 & $2.3\sigma$ &$\BFomBelle$&\\
\cline{1-7}
$ B\rightarrow(\rho/\omega)\gamma $ & & $\significance\sigma$ & $\BFav $\hspace*{0.5cm} & \hspace*{0.5cm}36.9&$5.1\sigma$ & $\BFavBelle$&\\
\cline{1-7}
$ B\rightarrow \rho\gamma $ & & $6.0\sigma$ & $\BFavrhorho $\hspace*{0.5cm}& & & & \\
\cline{1-7}
\end{tabular}
\end{table*}

The signal content of the data is determined by a multidimensional unbinned maximum likelihood 
fit, which is constructed individually for each of the three signal decay modes. All fits use 
\DeltaE , $\mes$,  $\cos\theta_H$,\footnote{
The helicity angle $\theta_H$ is defined as the angle between the $B$ momentum vector
and the $\pi^-$ track calculated in the $\rho$ rest frame in the case of a $\rho$ meson,
or the angle between the $B$ momentum vector and the normal to the $\omega$ decay plane
for an $\omega$ meson.}
and the neural network output $N$. For decays $\bomg$ ($\omega\to\pi^+\pi^-\pi^0$), the cosine 
of the angle between the \pip and \piz momenta in the $\pi^+\pi^-$ rest frame (Dalitz angle) 
is added as a fifth observable. In the fit, signal, continuum background, $\bkg$ decays, and 
other $B$ backgrounds are considered as hypotheses for the origin of the events.

Figure \ref{fig:rhoChfit} shows the data points and the projections of the fit results for 
\DeltaE and  $\mes$  separately for each decay mode. The  signal yields and corresponding 
branching fractions are listed in Table~\ref{tab:rgresults}. 

The isospin symmetry is tested by measuring the quantity 
$\Gamma(\brpg)/[2\Gamma(\brzg)] - 1 = \IST $, which -- within the large 
uncertainties -- agrees with the theoretical expectation~\cite{ali2004}.
The measured branching fractions are consistent with 
the isospin relation $ \Gamma_{B\to\rho^+\gamma}=2\Gamma_{B\to\rho^0\gamma}=2\Gamma_{B\to\omega \gamma}$ 
among the widths of the individual decays. Imposing this relation as a constraint, 
the isospin-averaged branching fraction is determined from a simultaneous
fit to the three decay modes to be
\begin{equation}\label{eq:res1}
  \avbr= (\BFav) \times10^{-6}\, ;
\end{equation}
the significance of the signal is \significance $\sigma$, including
systematic uncertainties. Excluding the $\Bz\to\omega\gamma$ mode from 
the simultaneous fit yields $\BR(B\to\rho\gamma) = (\BFavrhorho )\times10^{-6}$.

\begin{figure}[!h]
\begin{center}
\hspace*{-3.5mm}\includegraphics[width=270pt]{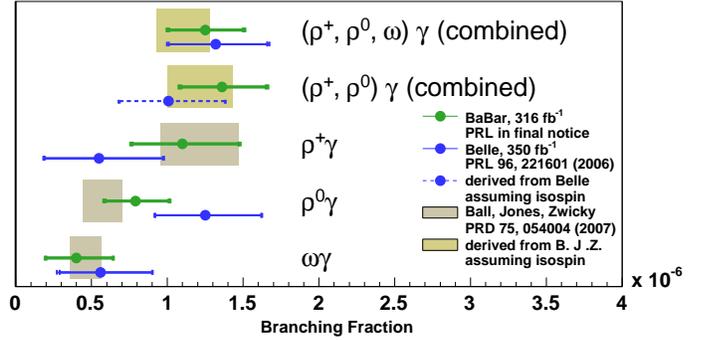}
\caption{Measured $B\to\rho\gamma$ and $B\to\omega\gamma$ branching fraction, compared to 
theoretical predictions. }
\label{fig:rgcomp} 
\end{center}
\end{figure}

As shown in Table~\ref{tab:rgresults} and Figure~\ref{fig:rgcomp}, all these results agree well with corresponding 
measurements by the Belle collaboration, published in~\cite{newbelle}.\footnote{
It should be noted that since this conference Belle has reported updated results based on 
a significantly larger dataset ($\approx600\ifb$). Good agreement with the previous measurements 
is found; still no significant signal is seen for the decay $\Bz\to\omega\gamma$. Details can 
be found in \cite{lp07belle}.}

Using the world average  value of $\BR(\bkg)$, Equation (\ref{eq:vtd}), and theory input from 
\cite{Ball:2006nr}, this translates into 
\begin{equation}\label{eq:res5}
\VtdVts = \VtdVtsval ,
\end{equation}
where the first error is experimental and the second is theoretical.
This result is in very good agreement with the
 measurement of this ratio from the study of $B^0$ and $B^0_s$ mixing~\cite{bsmixing}.

\begin{acknowledgments}
Thanks to the FPCP2007 organizers for making this productive and enjoyable conference possible. 
\end{acknowledgments}

\bigskip 


\begin{thebibliography}{99}   
\bibitem{bsm}
See, for example,
S.~Bertolini, F.~Borzumati, and A.~Masiero, \npb{294}, 321 (1987);
H.~Baer and  M.~Brhlik, \jprd{55}, 3201 (1997);
J.~Hewett and J.~Wells, \jprd{55}, 5549 (1997);
M.~Carena {\it et al.}, \plb{499}, 141 (2001).
%
\bibitem{Kagan:1998ym}
A.~L.~Kagan and M.~Neubert, \epjc{7}, 5 (1999).
\bibitem{Buchmuller:2005zv}
O.~Buchm\"uller and H.~Fl\"acher, \jprd{73}, 073008 (2006).
\bibitem{Chen:2001fj}
S.~Chen {\it et al.} [CLEO Collaboration], \jprl{87}, 251807 (2001).
\bibitem{Koppenburg:2004fz}
P.~Koppenburg {\it et al.} [Belle Collaboration], \jprl{93}, 061803 (2004).
\bibitem{BABARSEMI}
B.~Aubert {\it et al.} [\babar\ Collaboration], \jprd{72}, 052004 (2005).
\bibitem{BABARINCL}
B.~Aubert {\it et al.} [\babar\ Collaboration], \jprl{97}, 171803 (2006).
\bibitem{Aubert:2003zw}
B.~Aubert {\it et al.} [\babar\ Collaboration], \jprl{92}, 071802 (2004).
\bibitem{Misiak:2006zs}
M.~Misiak {\it et al.}, \jprl{98}, 022002 (2007).
\bibitem{Becher:2006qw}
T.~Becher and M.~Neubert, \plb{637}, 251 (2006).
\bibitem{Andersen:2006hr}
J.~Andersen and E.~Gardi, JHEP {\bf 01}, 029 (2007).
\bibitem{hfag2006}
Heavy Flavor Averaging Group, hep-ex/0603003. 
\bibitem{oldlambda}
Y.~J.~Lee {\it et al.} [Belle Collaboration], 
Phys.\ Rev.\ Lett.\  {\bf 95}, 061802 (2005).
\bibitem{Cheng}
  H.~Y.~Cheng and K.~C.~Yang,
  Phys.\ Lett.\  B {\bf 533}, 271 (2002).
\bibitem{Geng}
  C.~Q.~Geng and Y.~K.~Hsiao,
  Phys.\ Lett.\  B {\bf 610}, 67 (2005).
\bibitem{newlambda}
  M.~Z.~Wang {\it et al.}  [Belle Collaboration],
  Phys.\ Rev.\  D {\bf 76}, 052004 (2007).
\bibitem{ali2004}
A.~Ali, E.~Lunghi, and A. Y. Parkhomenko, \plb{595}, 323 (2004).
%
\bibitem{SM}
S.~W.~Bosch and G.~Buchalla, \npb{621}, 459 (2002).
%
\bibitem{alivtdvtstheory}
A.~Ali and A.~Y.~Parkhomenko, \epjc{23}, 89 (2002).
%
\bibitem{bsmixing}
A.~Abulencia {\it et al.}, 
Phys.\ Rev.\ Lett.\  {\bf 97}, 242003 (2006).
\bibitem{oldbabar}
B.~Aubert {\it et al.} [\babar\ Collaboration],
\jprl{94}, 011801 (2005).
%
\bibitem{cleo}
T.~E.~Coan {\it et al.} [CLEO Collaboration],
\jprl{84}, 5283 (2000).
%
\bibitem{newbelle}
D.~Mohapatra {\it et al.}, [Belle Collaboration],
  \jprl{96}, 221601 (2006).
%
\bibitem{newbabar}
B.~Aubert {\it et al.} [\babar\ Collaboration],
\jprl{98}, 151802 (2007).
%
\bibitem{babartag}
B.~Aubert {\it et al.}, 
\jprl{89}, 201802 (2002).
%
\bibitem{lp07belle}
B.~Nakao, {\it Probing New Physics with rare B decays}, presentation at Lepton-Photon 2007,  
http://chep.knu.ac.kr/lp07/htm/S7/S07\_21.pdf.
%
\bibitem{Ball:2006nr}
 P.~Ball and R.~Zwicky,   JHEP {\bf 0604}, 046 (2006);  
 P.~Ball, G.~Jones, R.~Zwicky, \jprd{75}, 054004 (2007). 
\bibitem{Bosch:2004nd}
  S.~W.~Bosch and G.~Buchalla,  
  JHEP {\bf 0501}, 035 (2005). 
%
\end{thebibliography}
\end{document}
%